 \documentclass[aps, onecolumn, showpacs, amsmath]{revtex4-2}
\usepackage{dcolumn}
\usepackage{bm}
\usepackage{graphicx}
\usepackage{color}
\usepackage{hyperref}
\usepackage{amsthm}
\usepackage{amssymb}

\begin{document}
\def\be{\begin{equation}}
\def\ee{\end{equation}}

\def\bc{\begin{center}}
\def\ec{\end{center}}
\def\bea{\begin{eqnarray}}
\def\eea{\end{eqnarray}}
\newcommand{\avg}[1]{\langle{#1}\rangle}
\newcommand{\Avg}[1]{\left\langle{#1}\right\rangle}

\def\ie{\textit{i.e.}}
\def\etal{\textit{et al.}}
\def\m{\vec{m}}
\def\G{\mathcal{G}}

\newcommand{\davide}[1]{{\bf\color{blue}#1}}
\newcommand{\gin}[1]{{\bf\color{green}#1}}

\title{Short-time large deviations of first-passage functionals for high-order stochastic processes}
\author{Lulu Tian}
\author{Hanshuang Chen}\email{chenhshf@ahu.edu.cn}
\author{Guofeng Li}
\affiliation{School of Physics and Optoelectronic Engineering, Anhui University, Hefei 230601, China}
\begin{abstract}
We consider high-order stochastic processes $x(t)$ described by the Langevin equation $\frac{{{d^m}x\left( t \right)}}{{d{t^m}}}= \sqrt{2D} \xi(t)$, where $\xi(t)$ is a delta-correlated Gaussian noise with zero mean, and $D$ is the strength of noise. We focus on the short-time statistics of the first-passage functionals $A=\int_{0}^{T} \left[ x(t)\right] ^n dt$ along the trajectories starting from $x(0)=L$ and terminating whenever passing through the origin for the first-time at $t=T$. Using the optimal fluctuation method, we analytically obtain the most likely realizations of the first-passage processes for a given constraint $A$ with $n=0$ and 1, corresponding to the first-passage time itself and the area swept by the first-passage trajectory, respectively. 
The tail of the distribution of $A$ shows an essential singularity at $A \to 0$, $P_{m,n}(A |L) \sim \exp\left(-\frac{\alpha_{m,n}L^{2mn-n+2}}{D A^{2m-1}} \right)$, where the explicit expressions for the exponents $\alpha_{m,0}$ and $\alpha_{m,1}$ for arbitrary $m$ are obtained. 
\end{abstract}

\maketitle
\section{Introduction}
From fluctuation–dissipation relation in equilibrium systems \cite{zwanzig2001nonequilibrium} to fluctuation theorem in nonequilibrium systems \cite{seifert2012stochastic}, Brownian motion plays a central role in statistical mechanics and other fields.  As one of the simplest stochastic processes, Brownian motion is described by the Langevin equation
\begin{eqnarray}\label{eq0.0}
\frac{dx}{dt}= \sqrt{2D} \xi(t),
\end{eqnarray} 
where $\xi(t)$ is the Gaussian white noise with  $\langle {\xi (t)} \rangle  = 0$ and $\langle {\xi (t)}{\xi (t')} \rangle  = \delta ( {t - t'} )$, and $D$ is the strength of noise. Considering a Brownian trajectory with a fixed duration $t$, a functional can be constructed along the trajectory, 
\begin{eqnarray}\label{eq0.1}
A=\int_{0}^{t} Z\left[ x(\tau) \right] d\tau,
\end{eqnarray} 
where $Z(x)$ can be an arbitrary function of $x$. Due to the randomness of Brownian motion, the functional $A$ is a well-defined random variable, and thus one of main concerns is statistical properties of the functional $A$.

Brownian functionals appear across various disciplines including physics, chemistry, biology, computer science and mathematics \cite{majumdar2007brownian}. Since the celebrated Feynman–Kac formula \cite{kac1949distributions}, the statistical properties of the Brownian functionals with a fixed time interval in one dimension have been studied extensively. Two commonly studied Brownian functionals in the literature are the local time ($Z(x)=\delta(x-x_\ell)$) and residence or occupation time ($Z(x)=\mathbb{I}_{\left[ a, b \right] }(x)$, where $\mathbb{I}_{\left[ a, b \right] }(x)=1$ for $x \in \left[ a, b \right] $ and zero otherwise), which measures the time the particle spent in the vicinity of a given location $x_\ell$ and the time spent a given spatial range $\left[ a, b \right]$, respectively. These functionals have found applications in chemical reactions, catalytic processes, and fluorescent imaging in molecular biology \cite{agmon1984residence,PhysRevE.57.3937,nguyen2010spectral,PhysRevLett.121.090601,agmon2011single}. For one-dimensional Brownian motions starting from the origin, the cumulative distribution of the residence time spent on a half-axis was known as the celebrated arcsine law of L\'evy \cite{Levy1940ArcsineLaw,feller1971introduction,majumdar2007brownian}. Recent studies led to many extensions with different scenarios, such as diffusion under confinement \cite{PhysRevE.76.041139,PhysRevE.105.034132,kay2023extreme,huang2024extremal}, in the presence of a potential landscape \cite{PhysRevLett.89.060601,PhysRevE.73.051102}, heterogeneous diffusion processes \cite{PhysRevE.105.024113}, random acceleration process \cite{majumdar2010time,boutcheng2016occupation,reymbaut2011convex}, fractional Brownian motion \cite{sadhu2018generalized,sadhu2021functionals}, run-and-tumble motion \cite{PhysRevE.103.042119,SinghArcsinelaws_RTP}, resetting stochastic processes \cite{pal2019local,PhysRevE.103.052119,den2019properties,bressloff2020occupation,yan2023breakdown},  and multi-particles systems \cite{godreche2001statistics,PhysRevLett.107.170601,PhysRevE.108.064113,PhysRevE.109.044150,smith2024macroscopic}. Another form of functionals is $Z(x)=x^n$, where $n$ is a real number, and particularly $n=1$ corresponds to the area swept by the Brownian motion. Area distribution under a Brownian excursion or a Bessel excursion over a fixed time interval were also studied and found diverse applications in physics \cite{majumdar2004exact,PhysRevX.4.021036}.

In recent years, the study of first-passage Brownian functionals have received growing attention \cite{majumdar2007brownian}. One main reason is that first-passage problem underlies many important processes in nature, such as diffusion-limited reactions, neural firing, and animal foraging, etc  \cite{redner2001guide,bray2013persistence}. In this context, the duration is defined as the time of Brownian motion from an initial position to a prescribed position, e.g. the origin, for the first time. Different from the functionals with a fixed time interval, the first-passage time is not fixed but random. In the literature, a commonly employed functionals along a first-passage trajectory is $Z(x)=x^n$, see also the definition in Eq.(\ref{eq1.1}), where $n=0$ simply represents the first-passage time itself, and $n=1$ corresponds to the area swept by the first-passage trajectory. The first-passage area has been investigated in the context of Brownian motion \cite{kearney2005area,kearney2016normalized,majumdar2020statistics}, drifted Brownian motion \cite{kearney2007first,abundo2017joint}, in the presence of a harmonic potential (Ornstein-Uhlenbeck process) \cite{kearney2021statistics,abundo2023first} or a logarithmic potential (Bessel process) \cite{PhysRevE.108.044151}, and stochastic resetting \cite{JPA2022.55.234001,dubey2023first,abundo2023first2}.


Although encouraging progress has been made, in most of cases, however, the exact results for the distributions of Brownian functionals or first-passage Brownian functionals are unavailable. Fortunately, one can resort the large deviation theory to obtain desirable results in the long-time or short-time limit \cite{agranov2020airy}. Large deviation theory is beyond the central limit theorem, which quantifies not only the Gaussian fluctuations about the typical value, but also the relative likelihood of atypical fluctuations \cite{touchette2009large}. The latter is of particular interest since atypical fluctuations can lead to many striking effects on the underlying dynamical systems, such as escape from a metastable state \cite{WKBReview1}, dynamical phase transition \cite{nyawo2017minimal,PhysRevE.98.052103,PhysRevE.107.064133,PhysRevE.110.024107}, and anomalous scaling \cite{nickelsen2018anomalous,PhysRevE.100.042135,PhysRevE.105.014120,PhysRevE.105.064102,smith2024anomalous}. In the long-time limit, the large deviations can be determined by the rate function, which is the Legendre-Fenchel transform of the scaled cumulant generating function of time-averaged observables \cite{touchette2018introduction}. In the short-time limit (or the weak-noise limit), large deviations can accurately described by the optimal fluctuation method (OFM), which traces back to early literature \cite{halperin1966impurity,zittartz1966theory,lifshitz1968theory}. This OFM calculation is based on the path integral representations for stochastic systems subject to certain constraints, and then determine, via the variation of action functional, the optimal, i.e., the most likely path which possesses a minimum action. The OFM has found numerous applications in different areas of physics. When applied to Brownian motion, the OFM becomes geometrical optics \cite{meerson2019geometrical}. Recently, the OFM has been developed and applied in a number of studies of large deviations of Brownian motion with various constraints \cite{meerson2019large,agranov2020airy,meerson2020area,majumdar2020statistics,nickelsen2018anomalous,smith2024anomalous,bar2023geometrical} and other stochastic systems, such as fractional Brownian motion \cite{meerson2022geometrical,PhysRevE.109.014146} and random acceleration process \cite{meerson2023geometrical,chen2024short}, active Brownian motion \cite{majumdar2020toward} and others \cite{meerson2016large}.

In the present work, we consider high-order stochastic processes $x(t)$, introduced first in the mathematics and statistics literature \cite{shepp1966radon,wahba1978improper}, indexed by an integer $m \geq 1$, 
\begin{eqnarray}\label{eq1.0}
x^{(m)}(t)= \sqrt{2D} \xi(t),
\end{eqnarray} 
where the superscript $(m)$ denotes the $m$th derivative with respect to time. In particular, for $m=1$ the process $x(t)$ corresponds to standard Brownian motion, and for $m=2$ it represents the random acceleration process. Due the non-Markov properties of high-order processes, exact results for $m \geq 2$ are rare. The persistence properties of the processes for general $n$ were initiated in Ref.\cite{PhysRevLett.77.2867}, and were calculated approximately for small values of $m$ in Ref.\cite{PhysRevE.69.016106}.

\begin{figure}
	\centerline{\includegraphics*[width=0.5\columnwidth]{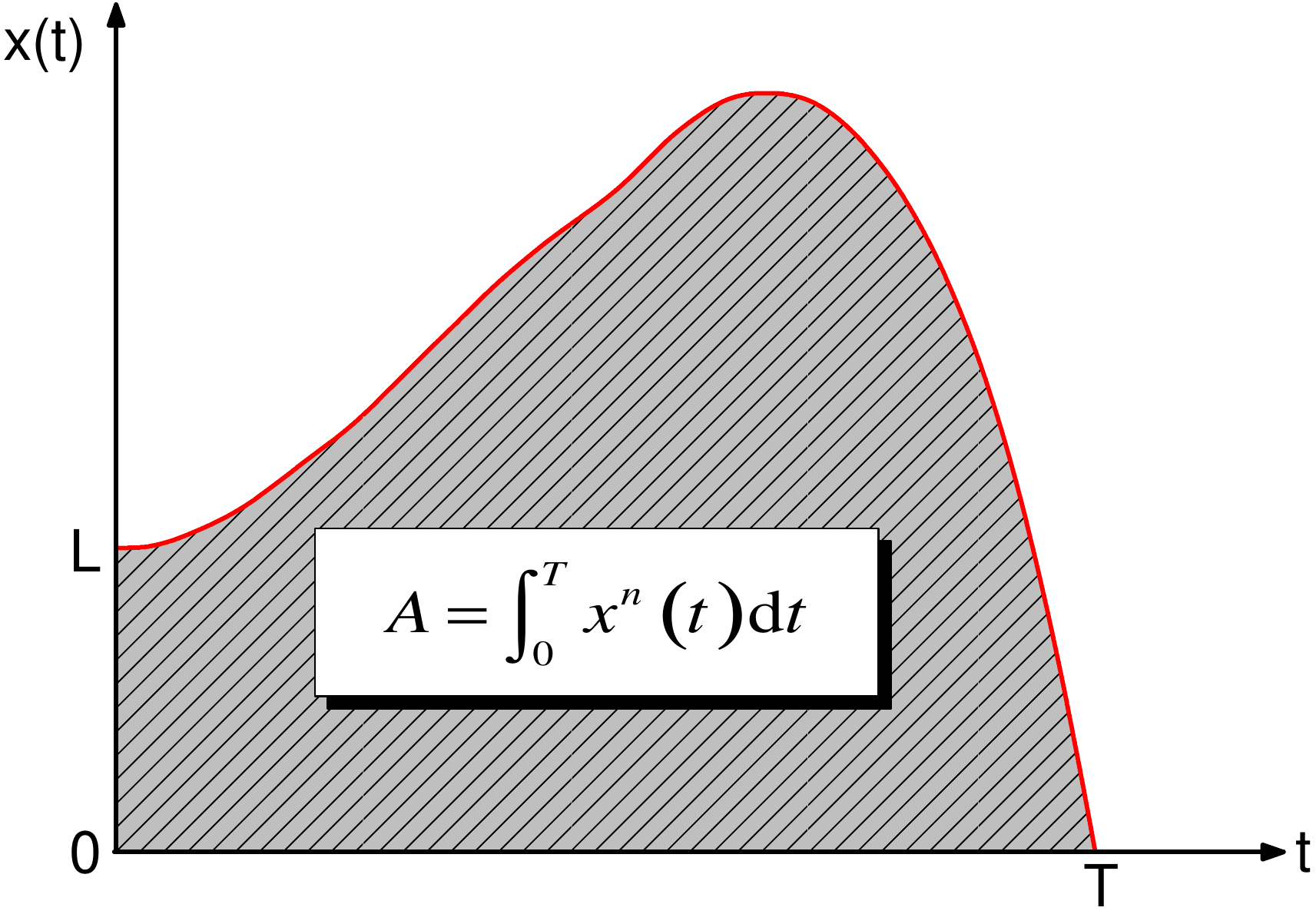}}
	\caption{First-passage trajectory $x(t)$ for the high-order process with $m=3$. The process starts from $x=L>0$ and terminates whenever $x(t)$ passes through the origin at a random time $t=T$. First-passage functionals along the stochastic trajectory is defined as ${A}=\int_{0}^{T}  x^n(t)  dt$.     \label{fig1}}
\end{figure}

We here focus on the short-time large deviations of the first-passage functionals $A$ of the high-order processes $x(t)$ with the form (see Fig.\ref{fig1} for an illustration)
\begin{eqnarray}\label{eq1.1}
{A}=\int_{0}^{T} \left[ x(t)\right] ^n dt,
\end{eqnarray} 
where $n \geq 0$ is an integer, and $T$ is the first-passage time of the process starting from the initial conditions
\begin{eqnarray}\label{eq1.6}
x(0)=L, \quad x'(0)=\cdots=x^{(m-1)}(0)=0,
\end{eqnarray}
and terminating whenever hitting the origin, i.e.,  
\begin{eqnarray}\label{eq1.6.1}
x(T)=0.
\end{eqnarray}

The main concern in the present work is placed on the $A \to 0$ tail of the probability distribution $P_{m,n}(A|L)$ of the functional $A$ for a set of given $m$, $n$ and $L$. Thanks to the OFM, the large-deviation tail can be obtained by the determination of the optimal path, that is the most likely realization of the process $x(t)$, conditioned on the specified value of $A \to 0$. The calculation shows that $P_{m,n}(A|L)$ has an essential singularity at $A \to 0$, $P_{m,n}(A |L) \sim \exp\left(-\frac{\alpha_{m,n}L^{2mn-n+2}}{D A^{2m-1}} \right)  $, where the exponent $\alpha_{m,n}$ is highly nontrivial, depending on the model parameters $m$ and $n$. We obtain analytically the nontrivial exponents $\alpha_{m,n}$ for $n=0, 1$ and for arbitrary $m$, see Eqs.(\ref{eq1.14}) and (\ref{eq2.7}), respectively. Our results are consistent with two recent works in Brownian motion ($m=1$) \cite{majumdar2020statistics} and random acceleration process ($m=2$) \cite{meerson2023geometrical}. Therefore, our work is an extension of the OFM in higher-order processes.

\section{Optimal fluctuation method for high-order processes}
The dimensional analysis suggests the following scaling behavior of $P_{m,n}(A|L)$,
\begin{eqnarray}\label{eq1.12}
P_{m,n}(A|L)=\frac{{{D^{1/\left( {2m - 1} \right)}}}}{{{L^{n + 2/\left( {2m - 1} \right)}}}}{F_{m,n}}\left( {\frac{{{D^{1/\left( {2m - 1} \right)}}A}}{{{L^{n + 2/\left( {2m - 1} \right)}}}}} \right),
\end{eqnarray} 
where $F_{m,n}(z)$ is a dimensionless scaling function of the dimensionless argument $z = {D^{1/\left( {2m - 1} \right)}}A/{L^{  n + 2/\left( {2m - 1} \right)}}$. To the best of our knowledge, the scaling function $F_{m,n}(z)$ is only known for $m=1$ where the process describes Brownian motion \cite{majumdar2020statistics}, which is a Markov process. While for $m \geq 2$, the dynamics is non-Markovian and the scaling function $F_{m,n}(z)$ is unknown at present. On the physical grounds $F_{m,n}(z)$ is expected to have a single maximum, of order 1, at $z \sim 1$, which indicates that the probability distribution $P_{m,n}(A|L)$ is expected to have its maximum, of order ${D^{1/\left( {2m - 1} \right)}}/{L^{  n + 2/\left( {2m - 1} \right)}}$, at $A \sim {L^{n + 2/\left( {2m - 1} \right)}}/{D^{1/\left( {2m - 1} \right)}}$. Instead of attempting to determine the entire scaling
function, in the present work we focus on the leading-order of $F_{m,n}(z)$ in the limit of $z \to 0$. This asymptotic corresponds to the $A \ll {L^{n + 2/\left( {2m - 1} \right)}}/{D^{1/\left( {2m - 1} \right)}}$ tail of the distribution $P_{m,n}(A|L)$.

Using the property of the Gaussian noise in the Langevin equation (\ref{eq1.0}), it is clear that the probability of any path $\left\lbrace x(t)\right\rbrace $ with $0\leq t \leq T$ can be written as  \cite{majumdar2007brownian}
\begin{eqnarray}\label{eq1.2}
P[\left\lbrace x(t)\right\rbrace] \propto  e^{-{S}},
\end{eqnarray} 
where  ${S}$ is called the action, given by
\begin{eqnarray}\label{eq1.3}
{S}=\frac{1}{4D} \int_{0}^{T}  \left[  x^{(m)} \right]  ^2  dt.
\end{eqnarray}

In the presence of constraints, the process is pushed into a large-deviation regime, and the action defined in Eq.(\ref{eq1.3}) becomes very large in the short-time limit (or equivalently in the weak-noise limit $D\to 0$). Therefore, the dominating contribution to the probability distribution comes from the optimal path: a single deterministic
trajectory $x^{*}(t)$ which minimizes the action functional Eq.(\ref{eq1.3})  subject to the specified additional constraints. The minimization procedure leads to the Euler-Lagrange equation for the optimal path. Once the optimal path
$x^{*}(t)$ is determined, one can evaluate the probability distribution of the specific large deviation up to a pre-exponential factor,
\begin{eqnarray}\label{eq1.3.1}
-\ln P \simeq S[x^{*}(t)],
\end{eqnarray} 
by plugging the optimal path into the action functional $S[x(t)]$ given by Eq.(\ref{eq1.3}).

We begin with a modified action functionals subject to the specified additional constraint Eq.(\ref{eq1.1}), 
\begin{eqnarray}\label{eq1.4}
\mathcal{S}_{\lambda}=\frac{1}{2D}\int_{0}^{T} \left[  \frac{1}{2} \left( x^{(m)} \right)^2 -\lambda x^n(t) \right]  dt,
\end{eqnarray}
where $\lambda$ is the Lagrange multiplier that enforces the constraint $\mathcal{A}=A$. The minimization of $\mathcal{S}_{\lambda}$ can be done via the variation of the constrained action functionals, which results in the Euler-Lagrange equation (see appendix \ref{app0} for details)
\begin{eqnarray}\label{eq1.5}
{x^{\left( 2m \right)}}\left( t \right)  - \lambda n {x^{n - 1}}\left( t \right) = 0.
\end{eqnarray}
To solve Eq.(\ref{eq1.5}), the boundary conditions Eq.(\ref{eq1.6}) and Eq.(\ref{eq1.6.1}) must be taken into account. However, these boundary conditions are not enough. The others can be obtained from the variation of action functionals (see appendix \ref{app0} for the derivation), 
\begin{eqnarray}\label{eq1.7}
\quad x^{(m)}(T)=\cdots=x^{(2m-2)}(T)=0.
\end{eqnarray}

In addition, the action should be minimized with respect to $T$, which leads to the optimal value of the first-passage time $T_*$. As shown in the appendix \ref{app0}, this additional minimization brings about an extra condition
\begin{eqnarray}\label{eq1.8}
x^{(2m-1)}(T_*)=0.
\end{eqnarray}

Once the optimal path $x^{*}(t)$ and the optimal first-passage time $T_*$ are found, we can express the Lagrange multiplier $\lambda$ in terms of $A$ from Eq.(\ref{eq1.1}). Then, using the action defined in Eq.(\ref{eq1.3}) we obtain, up to a preexponential factor, the $A \to 0$ tail of $P_{m,n}(A|L)$. It scales as
\begin{eqnarray}\label{eq1.16}
-\ln P_{m,n}(A|L) \simeq S\left[ x^{*}(t) \right]  =\frac{\alpha_{m,n}L^{2mn-n+2}}{D A^{2m-1}}.
\end{eqnarray}
Equation (\ref{eq1.16}) indicates that the distribution $P_{m,n}(A|L)$ has a universal essential singularity $ \sim \exp \left(-A^{1-2m}  \right)  $ for the $A \to 0$ tail. The exponents $\alpha_{m,n}$ are highly nontrivial depending on the order $m$ of the stochastic process and the constrained parameter $n$ on path observables. For $m=1$, corresponding to standard Brownian motion, the full distribution of $A$ were exactly derived by Majumdar and Meerson \cite{majumdar2020statistics} and the exponents $\alpha_{1,n}$ were known, which are $\alpha_{1,n}=1/{(n+2)^2}$ for all $n>-2$. For $m \geq 2$, the high-processes are non-Markov, and the exact expressions for the full distribution of $A$ are unknown. However, for $m=2$, which is the random acceleration process, Meerson has recently used the OFM to derive the first three exponents \cite{meerson2023geometrical}, $\alpha_{2,0}=3/4$, $\alpha_{2,1}=108/625$, $\alpha_{2,2}=(27/256) \tanh^4(\pi/2)=0.074 625 \cdots $. The main purpose of the present work is to deduce the exponents $\alpha_{m,n}$ for higher-order processes, i.e., for $m\geq 3$. Using the OFM, we will obtain analytically the exponents $\alpha_{m,n}$ for $n=1, 2$, and for arbitrary $m \geq 1$, given by Eqs.(\ref{eq1.14}) and (\ref{eq2.7}), respectively.

\section{Large deviation exponents for $A \to 0$ tail}

\subsection{$n=0$}
For $n=0$, the functional $\mathcal{A}=T$ in Eq.(\ref{eq1.1}) is the first-passage time itself. The Euler-Lagrange equation (\ref{eq1.5}) becomes very simple: ${x^{\left(2 m \right)}}\left( t \right)   = 0$, whose general solution reads
\begin{eqnarray}\label{eq1.9}
x(t)=c_0+c_1 t+c_2 t^2+\cdots+c_{2m-1} t^{2m-1}.
\end{eqnarray}
Using the intial conditions in Eq.(\ref{eq1.6}), we obtain 
\begin{eqnarray}\label{eq1.10}
c_0=L, \quad c_1=\cdots=c_{m-1}=0.
\end{eqnarray} 
Inserting further the boundary conditions in Eqs.(\ref{eq1.6.1}) and (\ref{eq1.7}) to Eq.(\ref{eq1.9}), we obtain the following linear equations,
\begin{eqnarray}\label{eqs1.1}
\left\{ \begin{gathered}
{T^m}{c_m} + {T^{m + 1}} {c_{m + 1}} +  \cdots  +{T^{2m - 2}} {c_{2m - 2}} + {T^{2m - 1}} {c_{2m - 1}} =  - L \hfill \\
m!{c_m} + \frac{{\left( {m + 1} \right)!}}{{1!}}{c_{m + 1}}T +  \cdots  + \frac{{\left( {2m - 2} \right)!}}{{\left( {m - 2} \right)!}}{c_{2m - 2}}{T^{m - 2}} + \frac{{\left( {2m - 1} \right)!}}{{\left( {m - 1} \right)!}}{c_{2m - 1}}{T^{m - 1}} = 0 \hfill \\
\vdots  \hfill \\
\left( {2m - 3} \right)!{c_{2m - 3}} + \frac{{\left( {2m - 2} \right)!}}{{1!}}{c_{2m - 2}}T + \frac{{\left( {2m - 1} \right)!}}{{2!}}{c_{2m - 1}}{T^2} = 0 \hfill \\
\left( {2m - 2} \right)!{c_{2m - 2}} + \frac{{\left( {2m - 1} \right)!}}{{1!}}{c_{2m - 1}}T = 0 \hfill \\ 
\end{gathered}  \right.
\end{eqnarray}
We first let $c_{2m-1}$ be a parameter, and rewrite the last $(m-1)$ equations in Eq.(\ref{eqs1.1}) in the matrix form, 
\begin{eqnarray}\label{eqs1.2}
\textbf{A} \textbf{c} =\textbf{b} 
\end{eqnarray}
where 
\begin{eqnarray}\label{eqs1.3}
\textbf{A} =\left[ {\begin{array}{*{20}{c}}
	1&{C_{m + 1}^1T}& \cdots &{C_{2m - 3}^{m - 3}{T^{m - 3}}}&{C_{2m - 2}^{m - 2}{T^{m - 2}}} \\ 
	0&1& \cdots &{C_{2m - 3}^{m - 4}{T^{m - 4}}}&{C_{2m - 2}^{m - 3}{T^{m - 3}}} \\ 
	\vdots & \vdots & \ddots & \vdots & \vdots  \\ 
	0&0& \cdots &1&{C_{2m - 2}^1T} \\ 
	0&0& \cdots &0&1 
	\end{array}} \right],
\end{eqnarray}
\begin{eqnarray}\label{eqs1.4}
\textbf{c}=\left[ {\begin{array}{*{20}{c}}
	{{c_m}} \\ 
	{{c_{m + 1}}} \\ 
	\vdots  \\ 
	{{c_{2m - 3}}} \\ 
	{{c_{2m - 2}}} 
	\end{array}} \right],
\end{eqnarray}
and
\begin{eqnarray}\label{eqs1.5}
\textbf{b}=  - {c_{2m - 1}}\left[ {\begin{array}{*{20}{c}}
	{C_{2m - 1}^{m - 1}{T^{m - 1}}} \\ 
	{C_{2m - 1}^{m - 2}{T^{m - 2}}} \\ 
	\vdots  \\ 
	{C_{2m - 1}^2{T^2}} \\ 
	{C_{2m - 1}^1T} 
	\end{array}} \right]
\end{eqnarray}
with $C_m^n = {{m!} \mathord{\left/
		{\vphantom {{m!} {\left[ {n!\left( {m - n} \right)!} \right]}}} \right.
		\kern-\nulldelimiterspace} {\left[ {n!\left( {m - n} \right)!} \right]}}$ is binominal coefficient. Here, the coefficient matrix $\textbf{A}$ is an upper triangular matrix and its inverse can be easily obtained, given by
\begin{eqnarray}\label{eqs1.6}
\textbf{A}^{-1} =\left[ {\begin{array}{*{20}{c}}
	1&{ - C_{m + 1}^1T}& \cdots &{{{\left( { - 1} \right)}^{m - 3}}C_{2m - 3}^{m - 3}{T^{m - 3}}}&{{{\left( { - 1} \right)}^{m - 2}}C_{2m - 2}^{m - 2}{T^{m - 2}}} \\ 
	0&1& \cdots &{{{\left( { - 1} \right)}^{m - 4}}C_{2m - 3}^{m - 4}{T^{m - 4}}}&{{{\left( { - 1} \right)}^{m - 3}}C_{2m - 2}^{m - 3}{T^{m - 3}}} \\ 
	\vdots & \vdots & \ddots & \vdots & \vdots  \\ 
	0&0& \cdots &1&{ - C_{2m - 2}^1T} \\ 
	0&0& \cdots &0&1 
	\end{array}} \right]
\end{eqnarray}
or  
\begin{eqnarray}\label{eqs1.7}
\left(  \textbf{A}^{-1} \right) _{ij}=\left\{ \begin{array}{lll} 0,    & {\rm{for}} \quad i>j,  \\
1,    & {\rm{for}} \quad i=j,  \\ {\left( { - 1} \right)^{j - i}}C_{m + j - 1}^{j - i}{T^{j - i}} &  {\rm{for}} \quad i<j.
\end{array}  \right. 
\end{eqnarray} 

It follows that $\textbf{c}=\textbf{A}^{-1} \textbf{b}$, 
\begin{eqnarray}\label{eqs1.8}
{c_{m + i - 1}} &=& \sum\limits_{j = 1}^{m - 1} \left(  \textbf{A}^{-1} \right) _{ij} {b_j} \nonumber \\ &=&  - {c_{2m - 1}}{T^{m - i}}\sum\limits_{j = i}^{m - 1} {{{\left( { - 1} \right)}^{j - i}}C_{m + j - 1}^{j - i}} C_{2m - 1}^{m - j} \nonumber \\ &=& {\left( { - 1} \right)^{m - i}}{T^{m - i}}C_{2m - 1}^{m - i}{c_{2m - 1}}, \quad i=1,\cdots,m
\end{eqnarray} 
Substituting Eq.(\ref{eqs1.8}) into the first equation of Eq.(\ref{eqs1.1}), we have
\begin{eqnarray}\label{eqs1.9}
- L &=& \sum\limits_{i = 1}^m {{c_{m + i - 1}}{T^{m + i - 1}}}  \nonumber \\ &=& {\left( { - 1} \right)^m}{c_{2m - 1}}{T^{2m - 1}}\sum\limits_{i = 1}^m {{\left( { - 1} \right)}^i}C_{2m - 1}^{m - i} \nonumber \\ &=&  {c_{2m - 1}}\frac{{{{\left( { - 1} \right)}^{m + 1}}mC_{2m - 1}^{m - 1}}}{{2m - 1}}{T^{2m - 1}}.
\end{eqnarray} 
From Eq.(\ref{eqs1.9}), we obtain
\begin{eqnarray}\label{eqs1.10}
{c_{2m - 1}} = \frac{{{{\left( { - 1} \right)}^m}\left( {2m - 1} \right)}}{{mC_{2m - 1}^{m - 1}}}\frac{L}{{{T^{2m - 1}}}}.
\end{eqnarray} 
Substituting Eq.(\ref{eqs1.10}) into Eq.(\ref{eqs1.8}), we obtain
\begin{eqnarray}\label{eqs1.11}
{c_{m + i - 1}} = {\left( { - 1} \right)^i}\frac{{\left( {2m - 1} \right)C_{2m - 1}^{m - i}}}{{mC_{2m - 1}^{m - 1}}}\frac{L}{{{T^{m + i - 1}}}}, \quad i=1,\cdots,m.
\end{eqnarray}

\begin{figure}
	\centerline{\includegraphics*[width=0.5\columnwidth]{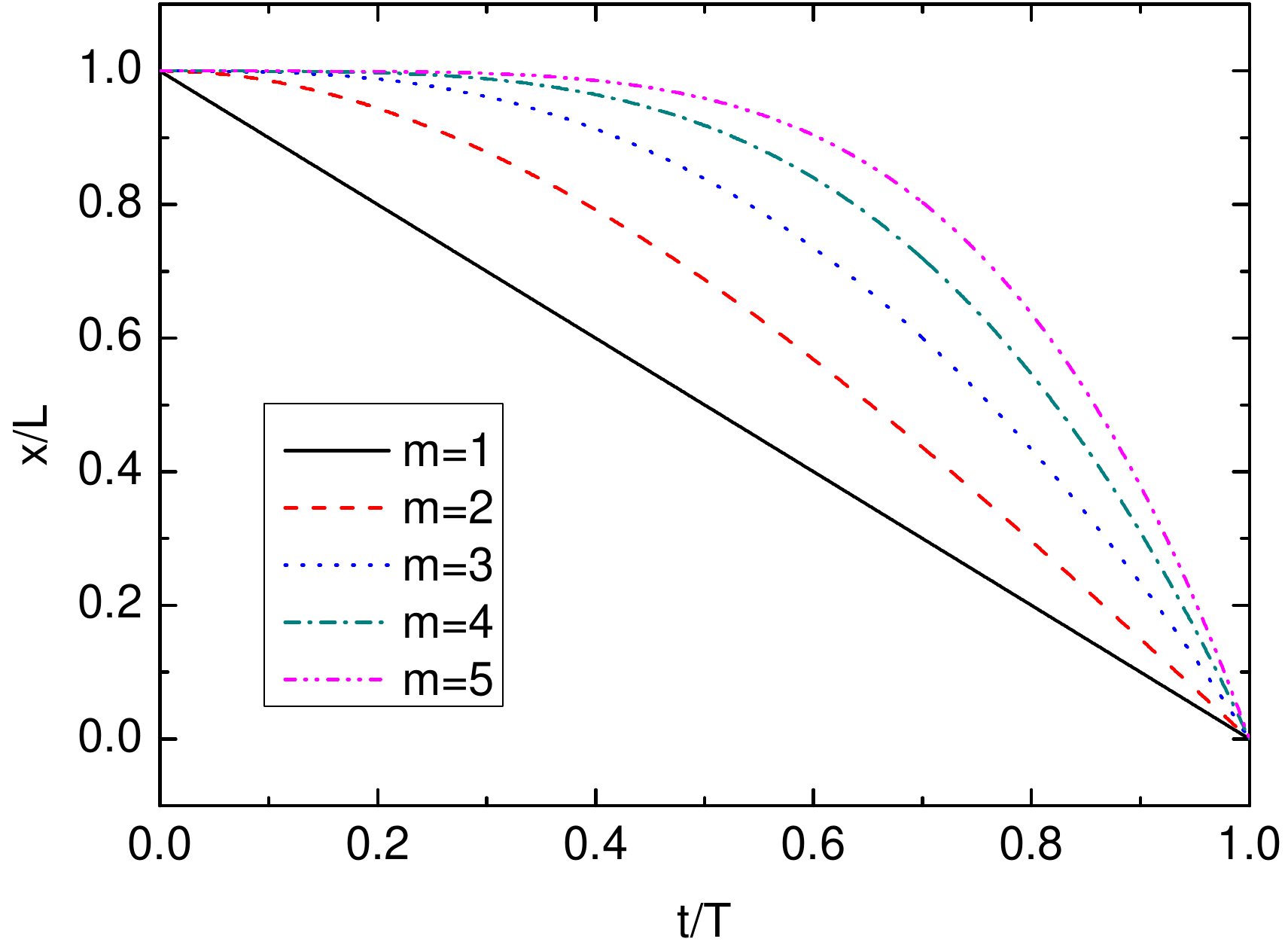}}
	\caption{Rescaled optimal path $x(t)/L$ as a function of $t/T$ for the high-order processes with the first five values of $m$, where the first-passage time $T$ is fixed.\label{fig2}}
\end{figure}

Using Eqs.(\ref{eq1.10}) and (\ref{eqs1.11}), we obtain the optimal path $x(t)$ for the high-order processes with a specified value of $T$,
\begin{eqnarray}\label{eq1.20}
\tilde x\left( {\tilde t} \right) = 1 + \sum\limits_{i = 1}^{m } {{{\tilde c}_{m + i - 1}}{{\tilde t}^{m + i - 1}}} ,
\end{eqnarray} 
where $\tilde x=x/L$, $\tilde t=t/T$, and ${\tilde c}_{m + i - 1}={c_{m + i - 1}} T^{m+i-1}/L={\left( { - 1} \right)^i}\left( {2m - 1} \right)C_{2m - 1}^{m - i}/\left( {m C_{2m - 1}^{m - 1}} \right)$. In Fig.\ref{fig2}, we show the optimal path $\tilde x(t)$ for the first five values of $m$. For $m=1$ (standard Brownian motion), the optimal path is a straight line, and thus the OFM becomes geometrical optics. For $m \geq 2$, the optimal path is no longer a straight line, and is now a polynomial of $t$ of order $2m-1$.

Using the optimal path in Eq.(\ref{eq1.20}), we calculate the action defined in Eq.(\ref{eq1.3}) 
\begin{eqnarray}\label{eq1.13}
-\ln P_{m,0}(T) \simeq S = \frac{{{\alpha _{m,0}}{L^2}}}{{D{T^{2m - 1}}}},
\end{eqnarray}
where the general expression of the exponent $\alpha_{m,0}$ is given explicitly
\begin{eqnarray}\label{eq1.14}
{\alpha _{m,0}}= \frac{{\left( {2m - 1} \right){{\left[ {\left( {m - 1} \right)!} \right]}^2}}}{4}.
\end{eqnarray}
For several small $m$, we have
\begin{eqnarray}\label{eq1.15}
{\alpha _{1,0}}=\frac{1}{4}, \quad {\alpha _{2,0}}=\frac{3}{4}, \quad {\alpha _{3,0}}=5, \quad {\alpha _{4,0}}=63, \quad {\alpha _{5,0}}=1296.
\end{eqnarray}
Obviously, for $m=1$ (Brownian motion) and $m=2$ (random acceleration process), our results are consistent with previous findings \cite{majumdar2020statistics,meerson2023geometrical}.

\subsection{$n=1$}
For $n=1$, the functional $\mathcal{A}=T$ defined in Eq.(\ref{eq1.1}) measures the area swept by the first-passage process. The Euler-Lagrange equation (\ref{eq1.5}) becomes: ${x^{\left(2 m \right)}}\left( t \right)   = \lambda$, whose solution is given by
\begin{eqnarray}\label{eq2.1}
x(t)=L+c_m t^m +\cdots+c_{2m-1} t^{2m-1}+\frac{\lambda}{(2m)!}t^{2m},
\end{eqnarray}
where we have used the boundary conditions in Eq.(\ref{eq1.6}). Substituting the boundary conditions in Eqs.(\ref{eq1.6.1}) and (\ref{eq1.7}) to Eq.(\ref{eq2.1}), we obtain the following linear equations,
\begin{eqnarray}\label{eqs2.1}
\left\{ \begin{gathered}
{T^m}{c_m} + {T^{m + 1}}{c_{m + 1}} +  \cdots  + {T^{2m - 2}}{c_{2m - 2}} + {T^{2m - 1}} {c_{2m - 1}} =  - L - \frac{\lambda }{{\left( {2m} \right)!}}{T^{2m}} \hfill \\
m!{c_m} + \frac{{\left( {m + 1} \right)!}}{{1!}}{c_{m + 1}}T +  \cdots  + \frac{{\left( {2m - 2} \right)!}}{{\left( {m - 2} \right)!}}{c_{2m - 2}}{T^{m - 2}} + \frac{{\left( {2m - 1} \right)!}}{{\left( {m - 1} \right)!}}{c_{2m - 1}}{T^{m - 1}} =  - \frac{\lambda }{{m!}}{T^m} \hfill \\
\vdots  \hfill \\
\left( {2m - 3} \right)!{c_{2m - 3}} + \frac{{\left( {2m - 2} \right)!}}{{1!}}{c_{2m - 2}}T + \frac{{\left( {2m - 1} \right)!}}{{2!}}{c_{2m - 1}}{T^2} =  - \frac{\lambda }{{3!}}{T^3} \hfill \\
\left( {2m - 2} \right)!{c_{2m - 2}} + \frac{{\left( {2m - 1} \right)!}}{{1!}}{c_{2m - 1}}T =  - \frac{\lambda }{{2!}}{T^2} \hfill \\ 
\end{gathered}  \right.
\end{eqnarray} 
As done in the last subsection, we first let $c_{2m-1}$ be a parameter, and rewrite the last $m-1$ equations in Eq.(\ref{eqs2.1}) in the matrix form, 
\begin{eqnarray}\label{eqs2.2}
\textbf{A} \textbf{c} =\textbf{d} ,
\end{eqnarray}
where $\textbf{A}$ and $\textbf{c}$ are given in Eq.(\ref{eqs1.3}) and Eq.(\ref{eqs1.4}), respectively. The vector $\textbf{d} $ in Eq.(\ref{eqs2.2}) is given by 
\begin{eqnarray}\label{eqs2.3}
\textbf{d} =\textbf{b}-\lambda \left[ {\begin{array}{*{20}{c}}
	{\frac{1}{{m!m!}}{T^m}} \\ 
	{\frac{1}{{\left( {m - 1} \right)\left( {m + 1} \right)!}}{T^{m - 1}}} \\ 
	\vdots  \\ 
	{\frac{1}{{3!\left( {2m - 3} \right)!}}{T^3}} \\ 
	{\frac{1}{{2!\left( {2m - 2} \right)!}}{T^2}} 
	\end{array}} \right],
\end{eqnarray}
where $\textbf{b}$ is given in Eq.(\ref{eqs1.5}). The solution of Eq.(\ref{eqs2.2}), $\textbf{c}=\textbf{A}^{-1} \textbf{d}$, reads
\begin{eqnarray}\label{eqs2.4}
{c_{m + i - 1}} = {\left( { - 1} \right)^{m - i}}{T^{m - i}}C_{2m - 1}^{m - i}{c_{2m - 1}} + {\left( { - 1} \right)^{m - i}}\frac{{\lambda {T^{m + 1 - i}}\left( {m - i} \right)}}{{\left( {m - i + 1} \right)!\left( {m + i - 1} \right)!}}, \quad i=1,\cdots,m.
\end{eqnarray}
Substituting Eq.(\ref{eqs2.4}) into the first line of Eq.(\ref{eqs2.1}), we obtain
\begin{eqnarray}\label{eqs2.5}
{c_{2m - 1}} = \frac{{{{\left( { - 1} \right)}^m}\left( {2m - 1} \right){{\left( {m!} \right)}^2}L + \lambda {T^{2m}}\left( {m - {m^2} - 1/2} \right)}}{{mC_{2m - 1}^{m - 1}{{\left( {m!} \right)}^2}{T^{2m - 1}}}}.
\end{eqnarray}
Using Eq.(\ref{eqs2.5}),  Eq.(\ref{eqs2.4}) can be rewritten as 
\begin{eqnarray}\label{eqs2.6}
{c_{m + i - 1}} = \frac{{{{\left( { - 1} \right)}^i}\left( {2m - 1} \right)C_{2m - 1}^{m - i}}}{{mC_{2m - 1}^{m - 1}{T^{m + i - 1}}}}\left[ {L + \frac{{{{\left( { - 1} \right)}^m}\lambda {T^{2m}}\left( {m - {m^2} - 1/2} \right)}}{{\left( {2m - 1} \right){{\left( {m!} \right)}^2}}}} \right] + \frac{{{{\left( { - 1} \right)}^{m - i}}\lambda {T^{m + 1 - i}}\left( {m - i} \right)}}{{\left( {m - i + 1} \right)!\left( {m + i - 1} \right)!}}, \quad i=1,\cdots,m. \nonumber \\
\end{eqnarray}

Using the constraint $A=\int_{0}^{T} x(t) dt$, we obtain the relation between $A$ and $\lambda$,
\begin{eqnarray}\label{eq2.3}
\lambda  = \frac{{{{\left( { - 1} \right)}^{m+1}}2\left( {2m + 1} \right){{\left( {m!} \right)}^2}}}{{{T^{2m + 1}}}}\left[ {\left( {2{m^2} + 1 - 2m} \right)LT - 2{m^2}A} \right].
\end{eqnarray}

The optimal value of first-passage time $T_*$ can be obtained from the boundary condition in Eq.(\ref{eq1.8}), which leads to 
\begin{eqnarray}\label{eq2.4}
\left( {2m - 1} \right)!{c_{2m - 1}} + \lambda T = 0.
\end{eqnarray}
Eq.(\ref{eq2.4}) can be simplified by using Eq.(\ref{eqs2.5}) and Eq.(\ref{eq2.3}), which yields the optimal first-passage time $T_*$ 
\begin{eqnarray}\label{eq2.4.1}
T_*=\frac{\kappa_mA}{L},
\end{eqnarray}
with
\begin{eqnarray}\label{eq2.4.4}
\kappa_m=\frac{2m+1}{2m-1}.
\end{eqnarray}

Utilizing Eq.(\ref{eq2.4.1}), Eq.(\ref{eq2.3})  and Eq.(\ref{eqs2.6}) can be respectively rewritten as
\begin{eqnarray}\label{eq2.4.2}
\lambda  =\frac{{2{{\left( { - 1} \right)}^{m + 1}}{{\left( {m!} \right)}^2}}}{{\kappa _m^{2m}}}\frac{{{L^{2m + 1}}}}{{{A^{2m}}}},
\end{eqnarray}
and
\begin{eqnarray}\label{eq2.4.3}
{c_{m + i - 1}} =\frac{{2{{\left( { - 1} \right)}^i}C_m^{m - i + 1}}}{{\kappa _m^{m + i - 1}C_{m + i - 1}^m}}  \frac{{{L^{m + i}}}}{{{A^{m + i - 1}}}}, \quad i=1,\cdots,m .
\end{eqnarray}
Using the rescaled coordinates, $\tilde{x}=x/L$ and $\tilde{t}=t/T^*$, the optimal path can be rewritten as
\begin{eqnarray}\label{eq2.20}
\tilde x\left( {\tilde t} \right) = 1 + \sum\limits_{i = 0}^m {{{\tilde c}_{m + i}}{{\tilde t}^{m + i}}} ,
\end{eqnarray}
where ${{{\tilde c}_{m + i}}}= c_{m+i} T_*^{m+i}/L$ is given by
\begin{eqnarray}\label{eq2.21}
{{{\tilde c}_{m + i}}}=  \frac{{2{{\left( { - 1} \right)}^{i + 1}}C_m^{m - i}}}{{C_{m + i}^m}},  \quad i=0,\cdots,m .
\end{eqnarray}

\begin{figure}
	\centerline{\includegraphics*[width=0.5\columnwidth]{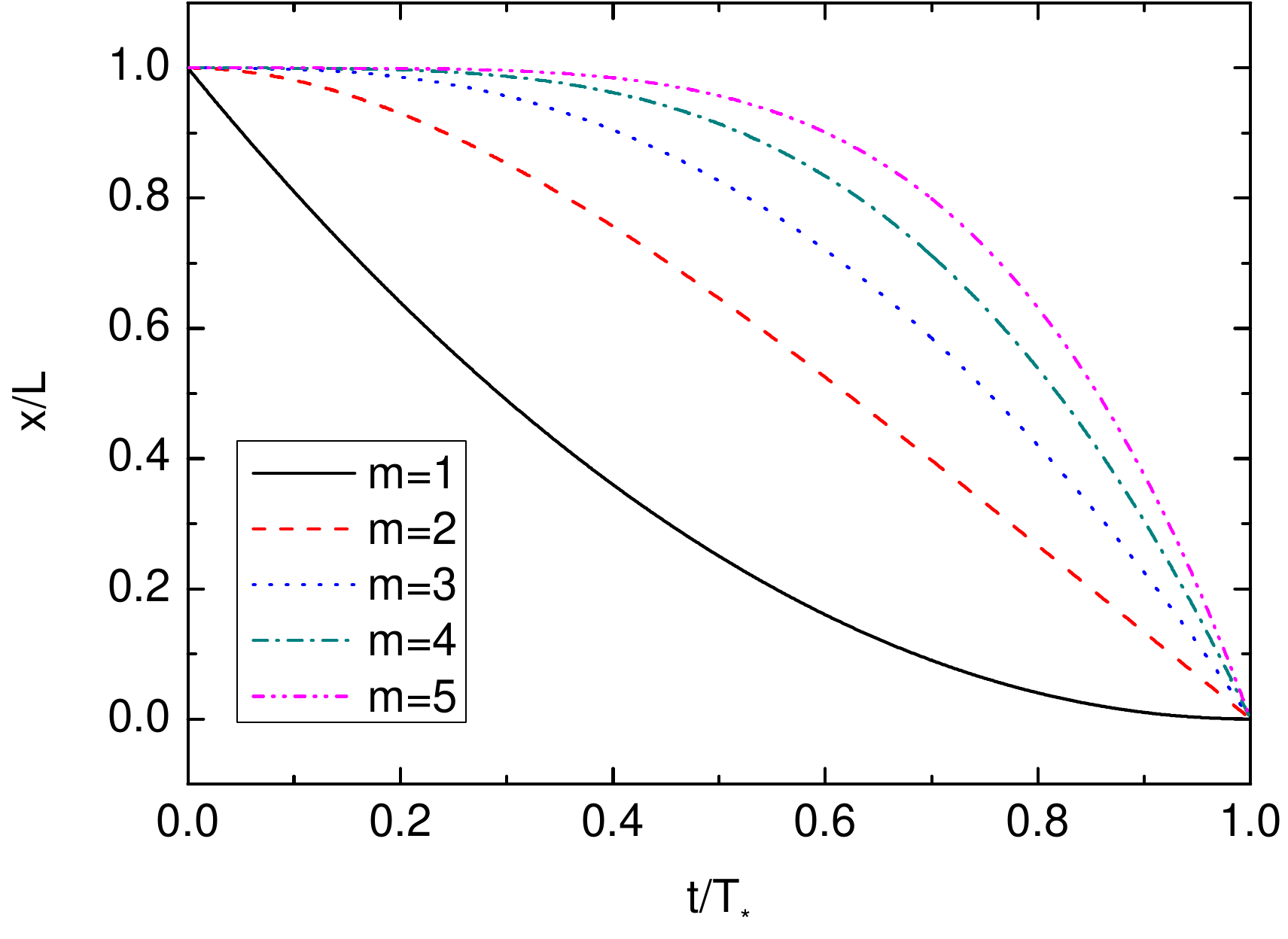}}
	\caption{Rescaled optimal path $x(t)/L$ as a function of $t/T_*$ for the high-order processes with the first five values of $m$, where the first-passage area $A=LT_*/\kappa_m$ is fixed with $\kappa_m=(2m+1)/(2m-1)$.\label{fig3}}
\end{figure}

In Fig.\ref{fig3}, we plot the optimal path in the rescaled coordinates for the first five values of $m$ with a specified first-passage area $A=LT_*/\kappa_m$.  The action in Eq.(\ref{eq1.3}) along the optimal path Eq.(\ref{eq2.20}) can be computed as
\begin{eqnarray}\label{eq2.5}
S &= & \frac{{{\lambda ^2}{T^{2m + 1}}}}{4D{{{\left( {m!} \right)}^2}\left( {2m + 1} \right)}}   + \frac{{\lambda }}{2D{m!}}\sum\limits_{i = 1}^m {\frac{{\left( {m + i - 1} \right)!}}{{\left( {i - 1} \right)!\left( {m + i} \right)}}{c_{m + i - 1}}{T^{m + i}}}  \nonumber \\&+& \frac{1}{4D}\sum\limits_{i = 1}^m {\sum\limits_{j = 1}^m {\frac{{\left( {m + i - 1} \right)!\left( {m + j - 1} \right)!}}{{\left( {i - 1} \right)!\left( {j - 1} \right)!\left( {i + j - 1} \right)}}{c_{m + i - 1}}{c_{m + j - 1}}{T^{i + j - 1}}} } 
\end{eqnarray}
Substituting Eq.(\ref{eq2.4.1}), Eq.(\ref{eq2.4.2}) and Eq.(\ref{eq2.4.3}) into Eq.(\ref{eq2.5}), we obtain
\begin{eqnarray}\label{eq2.6}
-\ln P_{m,1}(A) \simeq S = \frac{{{\alpha _{m,1}}{L^{2m+1}}}}{{D{A^{2m -1}}}},
\end{eqnarray}
where the exponent ${\alpha _{m,1}}$ is obtained explicitly
\begin{eqnarray}\label{eq2.7}
{\alpha _{m,1}}&=& \frac{{{{\left( {m!} \right)}^2}}}{{{\kappa_m ^{2m}}\left( {2m - 1} \right)}}=\frac{{{{\left( {m!} \right)}^2}{{\left( {2m - 1} \right)}^{2m - 1}}}}{{{{\left( {2m + 1} \right)}^{2m}}}}.
\end{eqnarray}
For several small $m$, we have
\begin{eqnarray}\label{eq2.8}
{\alpha _{1,1}}=\frac{1}{9}, \quad {\alpha _{2,1}}=\frac{108}{625}, \quad {\alpha _{3,1}}=\frac{112500}{117649}, \quad {\alpha _{4,1}}=\frac{52706752}{4782969}, \quad {\alpha _{5,1}}=\frac{5578855041600}{25937424601}.
\end{eqnarray}
One can compare the results in Eq.(\ref{eq2.8}) with previous works for $m=1$ (Brownian motion) and for $m=2$ (random acceleration process) \cite{majumdar2020statistics,meerson2023geometrical}, and find that they are exactly the same.

\section{Conclusions}
In conclusion, we have employed the OFM to study the short-time statistics of the first-passage functionals $A=\int_{0}^{T} x^n(t)dt$ for the high-order processes $x(t)$ indexed by a positive integer $m$. Via the variation of a modified action functional, we obtain the equation of motion for the most likely realization of $x(t)$. Combining boundary conditions, the equation of motion can be solved analytically for $n=0, 1$ and the optimal path $x(t)$ can be correspondingly obtained. The action along the optimal path can be also obtained analytically. As a result, the distribution of $A$ displays an essential singularity at the tail of $A \to 0$, $P_{m,n}(A |L) \sim \exp\left(-\frac{\alpha_{m,n}L^{2mn-n+2}}{D A^{2m-1}} \right)  $, where the exponent $\alpha_{m,n}$ is highly nontrivial. The main contribution in the present work is that we have obtained the explicit expressions of $\alpha_{m,n}$ for $n=0,1$ and for arbitrary $m$ (see Eqs.(\ref{eq1.14}) and (\ref{eq2.7})), which enables us to evaluate the exact $A\to 0$ tail of the distributions of the first-passage time itself and the area swept by the high-order processes till the first-passage time.

For $n=2$, the Euler-Lagrange equation (\ref{eq1.5}) can be analytically solved as well, but the solution is no longer a polynomial. This leads to the difficulty in obtaining the coefficients by the boundary conditions Eqs.(\ref{eq1.6}), (\ref{eq1.6.1}) and (\ref{eq1.7}), such that the determination of the exponents $\alpha_{m,2}$ for general $m$ is still a challenging task. However, for $m=1$ (Brownian motion) and $m=2$ (random acceleration process), the exponents $\alpha_{m,2}$ are known \cite{majumdar2020statistics,meerson2023geometrical}. For $n \ge 3$, the Euler-Lagrange equation (\ref{eq1.5}) cannot be solved analytically, and thus the exponents $\alpha_{m,n}$ for $n \ge 3$ can be only determined by numerical methods (except for $m=1$).

\appendix
\section{Derivations of Euler-Lagrange equation and boundary conditions}\label{app0}
We perform a linear variation of the constrained action functional defined in Eq.(\ref{eq1.4}),
\begin{eqnarray}\label{eqs0.1}
\delta {S_\lambda } = \int_0^T {\left( {{x^{\left( m \right)}}\delta {x^{\left( m \right)}} - \lambda n{x^{n - 1}}\delta x} \right)dt}  + \int_T^{T + \delta T} {\left( {\frac{1}{2}{{\left[ {{x^{\left( m \right)}}} \right]}^2} - \lambda {x^n}} \right)dt} .
\end{eqnarray}
Performing $m$ times integrations in parts in the first integral in Eq.(\ref{eqs0.1}), and evaluating the second integral in the limit of $\delta T \to 0$, we obtain
\begin{eqnarray}\label{eqs0.2}
\delta {S_\lambda } &=&   {\left( { - 1} \right)^m}\int_0^T {\left( {{x^{\left( {2m} \right)}} - \lambda n{x^{n - 1}}} \right)\delta xdt}  + \left. {{x^{\left( m \right)}}\delta {x^{\left( {m - 1} \right)}}} \right|_0^T - \left. {{x^{\left( {m + 1} \right)}}\delta {x^{\left( {m - 2} \right)}}} \right|_0^T +  \cdots  + \left. {{x^{\left( {2m - 1} \right)}}\delta x} \right|_0^T  \nonumber \\&+& \left[ {\frac{1}{2}{{\left[ {{x^{\left( m \right)}}\left( T \right)} \right]}^2} - \lambda {x^n}\left( T \right) - {x^{\left( {2m - 1} \right)}}\left( T \right){x^{\left( {m - 1} \right)}}\left( T \right)} \right].
\end{eqnarray}
Each of the three terms in the variation must vanish independently for arbitrary $\delta x$ and $\delta T$. The first term yields the Euler-Lagrange equation $x^{(2m)}(t)-\lambda n x^{n-1}(t)=0$. According to boundary conditions at initial stage in Eq.(\ref{eq1.6}) and one at final stage $x(T)=0$, we obtain other boundary conditions at final stage: $x^{(m)}(T)=\cdots=x^{(2m-2)}(T)=0$. The last term leads to the condition: $x^{(2m-1)}(T)x^{(m-1)}(T)=0$. One has to select one of two options: $x^{(2m-1)}(T)=0$ or $x^{(m-1)}(T)$ (they cannot hold simultaneously, as the problem then would
be overdetermined). As one can check, the condition
$x^{(2m-1)}(T)=0$ would give a local maximum of the action functionals, whereas the condition $x^{(m-1)}(T)=0$ yields the desired minimum.

\begin{acknowledgments}     
	This work was supported by the National Natural Science Foundation of China (11875069), the Key Scientific Research Fund of Anhui Provincial Education Department (2023AH050116), and Anhui Project (Grant No. 2022AH020009)
\end{acknowledgments}


\end{document}